\documentclass[aps,prb,superscriptaddress,twocolumn,showpacs]{revtex4}

\usepackage{amsmath,amssymb}
\usepackage{graphicx}
\usepackage{feynmf}
\usepackage{color}

\begin{document}

\title{Quantum metrology with a transmon qutrit}

\date{\today}

\author{A.R.\ Shlyakhov}
\affiliation{Moscow Institute of Physics and Technology, Institutskii per.\ 9,
Dolgoprudny, 141700, Moscow District, Russia}
\author{V.V.\ Zemlyanov}
\affiliation{Moscow Institute of Physics and Technology, Institutskii per.\ 9,
Dolgoprudny, 141700, Moscow District, Russia}
\author{M.V.\ Suslov}
\affiliation{Moscow Institute of Physics and Technology, Institutskii per.\ 9,
Dolgoprudny, 141700, Moscow District, Russia}
\author{A.V.\ Lebedev}
\affiliation{Moscow Institute of Physics and Technology, Institutskii per.\ 9,
Dolgoprudny, 141700, Moscow District, Russia}
\affiliation{Theoretische Physik, Wolfgang-Pauli-Strasse 27, ETH Zurich,
CH-8093 Z\"urich, Switzerland}
\author{G.S.\ Paraoanu}
\affiliation{Low Temperature Laboratory and QTF Centre of Excellence, Department of Applied Physics,
Aalto University School of Science, PO Box 15100, Aalto FI-00076, Finland}
\author{G.B.\ Lesovik}
\affiliation{Moscow Institute of Physics and Technology, Institutskii per.\ 9,
Dolgoprudny, 141700, Moscow District, Russia}
\author{G.\ Blatter}
\affiliation{Theoretische Physik, Wolfgang-Pauli-Strasse 27, ETH Zurich,
CH-8093 Z\"urich, Switzerland}

\begin{abstract}
Making use of coherence and entanglement as metrological quantum resources
allows to improve the measurement precision from the shot-noise- or quantum
limit to the Heisenberg limit. Quantum metrology then relies on the
availability of quantum engineered systems that involve controllable quantum
degrees of freedom which are sensitive to the measured quantity. Sensors
operating in the qubit mode and exploiting their coherence in a
phase-sensitive measurement have been shown to approach the Heisenberg scaling
in precision. Here, we show that this result can be further improved by
operating the quantum sensor in the qudit mode, i.e., by exploiting $d$ rather
than 2 levels. Specifically, we describe the metrological algorithm for using
a superconducting transmon device operating in a qutrit mode as a
magnetometer.  The algorithm is based on the base-3 semi-quantum Fourier
transformation and enhances the quantum theoretical performance of the sensor
by a factor 2. Even more, the practical gain of our qutrit-implementation is
found in a reduction of the number of iteration steps of the quantum Fourier
transformation by a factor $\log 2/\log 3 \approx 0.63$ as compared to the
qubit mode. We show, that a two-tone capacitively coupled rf-signal is
sufficient for the implementation of the algorithm.
\end{abstract}


\maketitle

The idea to boost metrological precision with the help of quantum resources
underwent an impressive development during recent years
\cite{Giovanetti:2011,Degen:2017}. Both types of quantum resources, coherence
and entanglement, are used in either sequential or parallel strategies,
respectively \cite{Giovanetti:2006}.  A key role in this endeavor is played by
novel quantum algorithms, in particular, Kitaev's phase estimation
\cite{Kitaev:1995,Cleve:1998} or the quantum Fourier transformation and its
semi-classical variant \cite{Griffiths:1996}, both exploiting phase coherence
as their quantum resource and thus following the sequential strategy. Previous
theoretical and experimental work in this direction has all routed in a base-2
computational scheme that exploits qubits as measuring devices. In this paper,
we demonstrate that a superconducting transmon device\cite{Koch:2007} operated
in a qutrit (or base-3) mode offers an enhanced performance as a
magnetic-field sensor; we present a specific algorithm exploiting the
semi-classical quantum Fourier transform as well as the required radio
frequency (rf) voltage-pulses for its implementation.

The use of entanglement-free protocols in metrology has been developed in
several steps, first addressing the problem of measuring the magnitude of
classical fields \cite{Vaidman:2004}, followed by a suggestion to quantify the
mesoscopic magnetic field generated by an assembly of nuclear magnetic moments
\cite{Giedke:2006}, and a proposal to use nitrogen--vacancy (NV) centers in
diamond for nanoscale magnetometry \cite{Twamley:2011}.  The basic concept of
this type of measurement was first used in a measurement of an optical phase
through interferometry \cite{Higgins:2007} and followed by the implementation
of a high-dynamic-range magnetic-field sensor in the form of a NV center
\cite{Wrachtrup:2012}.  An alternative route has been taken by starting from
the statistical counting of charge in mesoscopic transport \cite{Levitov:1996}
that required to include the measurement apparatus into the analysis.
Originally, the latter has been described by a spin that interacts with the
charge-transporting lead within a Gedankenexperiment. This idea has later been
taken to a realistic setup with a measurement device in the form of a charge-
or flux- qubit  \cite{Hassler:2006,Lebedev:2016}.  Subsequently, the
statistical counting has been refined to an algorithm that counts the number
of charges traversing the lead by making full use of quantum engineering ideas
in combination with the semi-classical Fourier transform  \cite{Lesovik:2010}.
While all of the above work is based on qubits or base-2 counting, the concept
of quantum counting \cite{Lesovik:2010} suggests a natural generalization of
such a scheme to qudits or base-$d$ counting \cite{Suslov:2011}.  Here, we
propose an application of the qudit counting algorithm to a metrological
measurement scheme for a magnetometer that makes use of a superconducting
transmon device \cite{Koch:2007} operated in a qutrit mode. Making efficient
use of the larger Hilbert space of a transmon qutrit and its specific linear
energy level dependence on the measured magnetic flux then allows for a faster
acquisition of information and thereby a more efficient measurement.

The algorithmic use of quantum engineered devices requires a non-linear
spectrum in order to address the quantum states individually.  Superconducting
circuit devices with Josephson junctions in loop geometries \cite{You:2011}
can be viewed as artificial atoms; they exhibit energy spectra with unequal
level spacings that can be designed on demand. Moreover, the position of
energy levels in such devices is sensitive to the magnetic flux penetrating
the SQUID loop of the artificial atom. In particular, the ultra-high
sensitivity of the transition frequency of the flux qubit \cite{Mooij:1999}
allows for its use as an ultra-high sensitive magnetic flux sensor
\cite{Bal:2012}.  Unfortunately, the high sensitivity of flux qubits to
low-frequency noise \cite{Yoshihara:2006} reduces their coherence time, that
makes them unfavorable for the implementation of quantum metrological
procedures. On the other hand, the special design of the transmon atom
\cite{Koch:2007} renders this device insensitive to the background charge
noise, resulting in larger coherence times.  Recently, the Kitaev- and
Fourier-like phase estimation algorithms have been successfully implemented in
a dc magnetic-flux measurement with a transmon qubit \cite{Danilin:2017},
resulting in an algorithmically improved sensitivity at high dynamic range.

The spectrum of a transmon atom is characterized by a particularly simple form
that corresponds to a harmonic oscillator with a weak non-linearity.
Furthermore, the transmon-atom's spectrum has a linear magnetic-flux
dependence of the excited states with respect to the ground state (to leading
order in the small ratio $E_{\rm \scriptscriptstyle C}/ E_{\rm
\scriptscriptstyle J}$ of charge- and Josephson energies of the transmon
atom).  Such a linear flux-dependence allows one to exploit several excited
levels of the transmon atom and implement a Fourier phase-estimation algorithm
operating in the qudit regime. On the other hand, the charge degree of freedom
enters the spectrum in a non-linear way, allowing for the transmon's
manipulation via capacitively coupled rf-fields.  Such coherent manipulation
of a transmon atom operating in a qutrit regime has been demonstrated recently
in several works\cite{Abdulmalikov:2013,Kumar:2016}; below, we will show how
to exploit the combination of these features in the effective operation of the
transmon as a magnetometer approaching Heisenberg scaling over the coherence
time of the device.

Operating the transmon in the Heisenberg limit, i.e., attaining a measurement
precision that scales in the invested resources $R$ as $1/R$ rather than
$1/\sqrt{R}$, requires a suitable metrological algorithm, in our case, the
semi-classical Fourier transform. The basic step in this algorithm is a Ramsey
cycle of duration $\tau$ that accumulates a phase $\phi = \mu H \tau /\hbar$
involving the magnetic field $H$ to be measured; here, the magnetic moment
$\mu$ plays the role of a coupling constant. Subsequent Ramsey cycles with
incrementally reduced times $\tau$, readout, and use of the result in the next
step allows to implement the quantum Fourier transformation (QFT) that boosts
the device performance to make it attain the Heisenberg limit. This QFT is
usually implemented in a binary system exploiting qubit devices. The
implementation of this algorithm in a ternary system with qutrits improves the
performance by a factor of 2, i.e., using half the resources one arrives at
the same precision as with the binary implementation. This
information-theoretical result is further improved when considering the number
of steps required for the same precision. Such a performance criterium makes
sense as most of the time required for the measurement is invested in the
preparation and readout of the transmon that has to be repeated in every step.
It turns out that the number of steps is reduced by a factor $\ln(2)/\ln(3)
\approx 0.63$ when replacing the qubit-mode by a qutrit implementation, a
considerable speedup of the measurement.

In the following, we briefly recapitulate the binary metrological algorithm
(Sec.\ \ref{sec:qubit}) and then extend the discussion to the ternary system
that exploits qutrits, see Sec.\ \ref{sec:qutrit}. The comparison between the
two procedures in Sec.\ \ref{sec:comp} leads us to the performance results
discussed above. In Sec.\ \ref{sec:transmon}, we discuss the spectral
properties of the transmon device and find the specific rf-pulses that prepare
the qutrit for the measurement and for the readout: Essentially, these pulses
generate a basis change (and back) from the computational basis (where the
qutrit is operated) to the `counting' or measurement basis where the field
imposes its characteristic `rotation' of the qutrit that encodes the unknown
magnitude of the field. It turns out, that a two-tone pulse addressing the the
first and second excited states is sufficient to perform this task. In Sec.\
\ref{sec:conclusion}, we summarize and conclude our discussion.

\section{Qubit metrological procedure}\label{sec:qubit}

We start with a brief summary of the standard qubit-based metrological
procedure for the measurement of a constant magnetic field. The elementary
step in this measurement scheme is a Ramsey interference experiment where the
qubit (or equivalently a spin-1/2) is prepared in an equally weighted (or
balanced) superposition of its $z$-polarized states  $|\!\!\downarrow\rangle$
and $|\!\!\uparrow\rangle$, $|\psi_0\rangle = (|\!\!\downarrow\rangle
+|\!\!\uparrow\rangle)/\sqrt{2}$.  The state $|\psi_0\rangle$ is located in
the equatorial $xy$-plane of the Bloch sphere and can be prepared by
performing a $\pi/2$ rotation of the $|\!\!  \downarrow\rangle$ state around a
$y$-axis, $|\psi_0\rangle = \hat{U}_y(\pi/2) |\!\!\downarrow\rangle$. Next,
the state $|\psi_0\rangle$ is exposed during a time interval $\tau$ to a
magnetic field $H$ directed along the $z$-axis, $|\psi_0\rangle \to
\exp(-i\hat\sigma_z \phi/2) |\psi_0\rangle \equiv |\psi_\phi\rangle$, with
$\phi = \mu H \tau/\hbar$ and $\mu$ is the magnetic moment of the spin. As a
result, the initial state $|\psi_0\rangle$ picks up the additional
field-sensitive relative phase $\phi$, $|\psi_\phi\rangle = (e^{-i\phi/2}
|\!\!\downarrow\rangle + e^{i\phi/2} |\!\!\uparrow \rangle) /\sqrt{2}$. The
last step of the Ramsey interference experiment is the readout procedure,
where the information about the value of the magnetic field encoded in the
qubit state is extracted through a projective measurement of the spin
polarization along the $z$-axis.  In order to make the outcome probabilities
$P_{\uparrow, \downarrow}$ depend on the magnetic field $H$, the state
$|\psi_\phi\rangle$ is first transformed by applying a unitary readout
operation, which coincides with that for the preparation for the qubit case,
$|\psi_\mathrm{out}\rangle\! = \hat{U}_y(\pi/2) |\psi_\phi\rangle =
-i\sin(\phi/2) {|\!\!\downarrow\rangle} + \cos(\phi/2)|\!\!\uparrow\rangle$.
Repeating this Ramsey cycle several times, one can accumulate enough
statistics and extract the value of the field $H$, see Refs.\
[\onlinecite{Suslov:2011,Lebedev:2016}].

Quite importantly, for the specific situation where the magnetic field $H$ can
assume only two values $H = 0$ and $H = h$, one can unambiguously distinguish
between these two possibilities during a single Ramsey cycle, i.e., a
single-shot measurement.  Indeed, adjusting the time delay $\tau$ such that
$\phi = \mu h\tau/\hbar = \pi$, one has that either $|\psi_\phi\rangle =
|\psi_0\rangle$ or $|\psi_\phi\rangle = |\psi_1\rangle = (|\langle \downarrow
\rangle - |\uparrow\rangle)/\sqrt{2}$ and hence, $|\psi_\mathrm{out}\rangle =
|\uparrow\rangle$ or $|\psi_\mathrm{out}\rangle = |\downarrow\rangle$. This
results in the probabilities $P_\uparrow = 1$ and $P_\downarrow = 0$ for $H=0$
or $P_\uparrow = 0$ and $P_\downarrow = 1$ for $H = h$, allowing for a
single-shot distinction between the two field values.  The basis states
$|\uparrow \rangle$ and $|\downarrow \rangle$ then define the so-called
computational basis, while the states $|\psi_0\rangle$ and $|\psi_1\rangle$
form the counting basis.

The above remarkable fact can be further exploited to distinguish between
$2^K$ discrete magnetic-field values with only $K$ Ramsey experiments: Let the
magnetic field $H \in [0,2h_0]$ assume only discrete values that correspond to
an exact $K$-bit fractional binary representation of the form
\begin{equation}
   H = h_0\, \Bigl( \frac{b_0}{2^0} + \frac{b_1}{2^1} + \frac{b_2}{2^2} +
   \dots + \frac{b_{K-1}}{2^{K-1}} \Bigr),
   \label{eq:Hbinary}
\end{equation}
where the amplitudes $b_n$, $n=0,\dots,K-1$, take binary values $0$ and $1$.
Let us also choose an elementary time delay $\tau_0$ such that $\mu h_0
\tau_0/\hbar = \pi$. A Ramsey measurement with an enhanced time delay
$\tau_{K-1} = 2^{K-1} \tau_0$ then accumulates a phase $\phi_{K-1} = \pi
b_{K-1} +2\pi n$, where the integer $n= 2^{K-1}b_0+\dots+ 2b_{K-2}$ is given
by the previous bits $b_n$, $n=0,\dots,K-2$. Although this first Ramsey
experiment provides the phase $\phi_{K-1}$ only modulo $2\pi$, the even or odd
outcome for this modulo-$2\pi$ phase allows for an unambiguous identification
of the last binary digit $b_{K-1}$.

In the next step, the time delay in the Ramsey experiment is twice reduced,
$\tau_{K-2} = 2^{K-2}\tau_0$. The accumulated field-sensitive phase is now
given by $\phi_{K-2} = \pi (b_{K-2} + b_{K-1}/2) \mod 2\pi$. Since we have
already learned the value of the bit $b_{K-1}$ in the previous measurement, we
can apply an additional rotation $\hat{U}_z(-\pi b_{K-1}/2)$ prior to the
readout operation in order to compensate for the residual phase $\pi
b_{K-1}/2$. The subsequent readout operation and measurement along $z$
provides the next bit $b_{K-2}$ in a deterministic way. Proceeding analogously
with gradually decreased time delays $2^{K-3}\tau_0, 2^{K-4}\tau_0,\dots,
\tau_0$ allows for an unambiguous determination of all bits $b_{K-1},
b_{K-2},\dots, b_0$ in the binary representation of the magnetic field $H$ and
thus its precise selection out of the $2^K$ discrete allowed values, see Eq.\
\eqref{eq:Hbinary}.

\section{Qutrit metrology}\label{sec:qutrit}

Next, we consider the generalization of the above qubit-based metrological
scheme to a qutrit, i.e., a quantum system that is endowed with a
three-dimensional Hilbert space. Let the quantum system in question be a
spin-1 system with (computational) basis states $|0\rangle$, $|1\rangle$, and
$|2\rangle$ corresponding to the $m_z=-1, 0, 1$ angular-momentum polarizations
along the $z$-axis. As with the qubit case, we prepare the qutrit in a
\textit{balanced} state $|\psi_0\rangle = (|0\rangle +|1\rangle
+|2\rangle)/\sqrt{3}$ and expose it to a constant magnetic field $H$ directed
along $z$-axis during the time $\tau$. As a result, the information about the
value of the field is encoded into the relative phases of the qutrit state,
\begin{equation}
  |\psi_\phi\rangle = \frac1{\sqrt{3}}\Bigl(|0\rangle + e^{i\phi} |1\rangle
  + e^{2i\phi} |2\rangle \Bigr),
  \label{eq:exposedqtrit}
\end{equation}
where $\phi = \mu H\tau/\hbar$ and we have omitted the overall phase factor
$e^{-i\phi}$.

To start with, we consider a situation where the magnetic field assumes only
one of three values $H \in \{0,h,2h\}$, $h > 0$; the task then is to
unambiguously distinguish between these three alternatives via a single-shot
measurement of the state (\ref{eq:exposedqtrit}); such a one-shot
discrimination is indeed possible as was shown in Ref.\
[\onlinecite{Suslov:2011}] within the context of the quantum counting problem.
We expose the initial balanced state $|\psi_0\rangle$ during a specific time
interval $\tau_0$ to the field such that the phase $\phi = \mu h\tau_0/\hbar$
assumes the value $\phi = 2\pi/3$.  As a result, the qutrit ends up in one of
the counting states $|\psi_\phi \rangle =|\psi_0\rangle$, $|\psi_\phi \rangle
=|\psi_1\rangle = (|0\rangle + e^{2\pi\,i/3}|1\rangle +
e^{-2\pi\,i/3}|2\rangle)/\sqrt{3}$ or $|\psi_\phi \rangle =|\psi_2\rangle =
(|0\rangle + e^{-2\pi\,i/3}|1\rangle + e^{2\pi\,i/3}|2\rangle)/\sqrt{3}$,
depending on the discrete field values $0,h$, or $2h$. Applying a base-$d$
quantum inverse Fourier transformation $\hat{F}^{-1}_d$,
\begin{equation}
  \hat{F}_d^{-1}|n\rangle = \frac1{\sqrt{d}} \sum_{k=0}^{d-1}
  \mathrm{e}^{-2\pi i nk/d} \, |k\rangle,
  \label{eq:invFourier}
\end{equation}
with $d = 3$ to these counting states, one can check that the resulting state
$|\psi_\mathrm{out}\rangle = \hat{F}^{-1}_3 |\psi_\phi\rangle$ coincides with
one of the computational states $|0\rangle$, $|1\rangle$, or $|2\rangle$,
depending on the magnetic field $H$ taking the values $0$, $h$ or $2h$,
respectively.  Therefore, measuring the polarization of the resulting state
$|\psi_\mathrm{out}\rangle$ along $z$-axis allows for an unambiguous
distinction between the three possible values of the magnetic field.

Next, consider the situation where the magnetic field $H$ has an exact ternary
representation
\begin{equation}
   H = h_0\, \Bigl( \frac{t_0}{3^0}+ \frac{t_1}{3^1} + \dots
   + \frac{t_{K-1}}{3^{K-1}} \Bigr),
   \label{eq:Htrinary}
\end{equation}
where the amplitudes (or trits) $t_n$, $n=0,\dots,K-1$, can take only three
discrete values $0,1$, and $2$. Then, similarly to the qubit case, one can
successively determine all $K$ trinary digits starting from the least
significant digit $t_{K-1}$ within $K$ separate preparation--exposure--readout
steps. Indeed, in the first step, we prepare the qutrit in the balanced state
$|\psi_0\rangle$ and expose it to the magnetic field $H$ during the time
interval $\tau_{K-1} = 3^{K-1} \tau_0$, where $\tau_0$ is chosen to satisfy
the relation $\mu h_0\tau_0/\hbar = 2\pi/3$. Then, the magnetic-field
dependent phase $\phi_{K-1} = (2\pi/3)\,t_{K-1} \mod 2\pi$ can only take the
three values $0, 2\pi/3$, and $4\pi/3$, which can be unambiguously
distinguished by the inverse Fourier transform in the readout step described
above. Next, the digit $t_{K-2}$ is determined by reducing the exposure time
by one-third, $\tau_{K-2} = 3^{K-2} \, \tau_0$, that provides the
field-dependent phase $\phi_{K-2} = (2\pi/3) (t_{K-2} +t_{K-1}/3)$. Making use
of the digit $t_{K-1}$ found in the previous step, the residual phase $2\pi
t_{K-1}/9$ is compensated before readout through projection along $z$, that
leads to the next trinary digit or trit $t_{K-2}$, and so on.

In the most general situation, the magnetic field $H$ assumes continuous
values and has no exact finite representation as in Eqs.\  (\ref{eq:Hbinary}) or (\ref{eq:Htrinary}). Hence, the number of trinary (or binary) digits needed to describe it is infinite and one cannot measure the field $H$ exactly with a finite number of preparation--exposure--readout steps. Still, we can find an approximate value of the field by detecting the first $K$ digits of its numerical representation. E.g., using a base-3 representation with qutrits and applying the magnetic field $H$ during the time interval $\tau_{K-1}$ to the balanced state $|\psi_0\rangle$, the magnetic field induces the phase $\phi_{K-1} = (2\pi/3) t_{K-1} + 3^{K-1} \delta\phi$, where the residual phase $\delta\phi \in [0, \pi/3^K]$ as given by
\begin{equation}
   \delta\phi = \frac{2\pi}3\, \Bigl( \frac{t_K}{3^K}
   + \frac{t_{K+1}}{3^{K+1}}+ \dots \Bigr)
\end{equation}
is unknown and cannot be compensated any more. As a consequence, rather than definitive outcomes, we have to find the probabilities $P_k$ to observe the
qutrit in the states $|k\rangle$, $k=0,1,2$.  Applying the inverse Fourier
transform and analyzing the result, we find that the probabilities
\begin{eqnarray} \label{eq:P012}
  && P_0 = \frac19 \bigl[ 1 + 2\cos \bigl( {2\pi}\, t_{K-1}/3
           +3^{K-1}\delta\phi\bigr) \bigr]^2, \\
  && P_1 = \frac19 \bigl[ 1 + 2\cos \bigl( {2\pi}\, (t_{K-1}-1)/3
           +3^{K-1}\delta\phi \bigr) \bigl]^2, \nonumber \\
  && P_2 = \frac19 \bigl[ 1 + 2\cos \bigl( {2\pi}\, (t_{K-1}-2)/3
           + 3^{K-1}\delta\phi \bigr) \bigl]^2,  \nonumber
\end{eqnarray}
deviate from zero and unity due to the unknown phase $\delta \phi$, hence, we
cannot any more distinguish between different $t_{K-1}$ unambiguously.
Instead, we have to resort to a statistical analysis and select between the
three alternatives $t_{K-1} = 0,1$ or $2$ by finding the maximum  probability
$P_k$, $k=0,1,2$. In practice, the weights of the three probabilities in Eq.\
\eqref{eq:P012} are disjoint and a single measurement is sufficient to
determine the trit's value with good confidence; the overall success
probabilities for the measurement schemes discussed below are calculated on
the basis of this assumption.  Repeating the procedure for the remaining $K-1$
trits including the required phase compensation prior to the readout, one
arrives at a set of $K$ trinary digits $\vec{t} \equiv t_0,\dots, t_{K-1}$.
The overall probability to observe the trinary string $\vec{t}$, given some
unknown magnetic field $H$, is given by
\begin{eqnarray}
   P\bigl(\vec{t}\>|H) = \prod_{k=0}^{K-1}
   \frac19\, \bigl[ 1+2\cos\bigl(3^k (\phi(H) - \tilde\phi_{\vec{t}}) \bigr)\bigr]^2,
\end{eqnarray}
where the phase $\tilde\phi_{\vec{t}} = (2\pi/3)\sum_{k=0}^{K-1} t_k/3^k$
relates to the string $\vec{t}$ and $\phi(H) = \mu H \tau_0/\hbar$ is the true field-induced phase; for an exact trinary value of $H$ and its associated string $\vec{t}$, we have $P\bigl(\vec{t}\>|H) = 1$.

In a next step, we make use of Bayes' theorem and infer the probability
$P(\phi(H)|\vec{t}\>)$ for the accumulated phase $\phi(H)$, provided we have observed a string $\vec{t}$, $P(\phi(H)|\vec{t}\>) \propto P(\vec{t}\>|H)$.
Making use of the trigonometric identity $\sin(3\alpha) = \sin(\alpha)[3-4\sin^2(\alpha)]$, one finds that,
\begin{equation}
  P\bigl(\phi(H)|\vec{t}\>) = \frac1{2\pi}\,\frac{\sin^2[3^K (\phi(H) -
  \tilde\phi_{\vec{t}})/2]}{3^K\sin^2[(\phi(h)-\phi_{\vec{t}})/2]}.
  \label{eq:qutritprob}
\end{equation}
A similar analysis provides the result
\begin{equation}
  P\bigl(\phi(H)|\vec{b}\>) = \frac1{2\pi}\,\frac{\sin^2[2^K (\phi(H)
  -\tilde\phi_{\vec{b}})/2]}{2^K\sin^2[(\phi(H)-\phi_{\vec{b}})/2]}
   \label{eq:qubitprob}
\end{equation}
for the qubit-based protocol, where $\vec{b}$ is the $K$-bit string learned
during the $K$-step measurement process and $\tilde\phi_{\vec{b}} = \pi
\sum_{k=0}^{K-1} b_k/2^k$.

The above qubit and qutrit Fourier metrological schemes can be generalized to
a setup with qudits that are endowed with a $d$-dimensional Hilbert space. The
qudit then is prepared in the balanced state $|\psi_0\rangle = (1/\sqrt{d})
\sum_{j=0}^{d-1} |j\rangle$ and subsequently experiences a magnetic-phase
accumulation $|\psi_0\rangle \to |\psi_\phi\rangle = (1/\sqrt{d})
\sum_{j=0}^{d-1} e^{ij\phi}|j\rangle$, followed by a $d$-base inverse Fourier
readout measurement. The posterior probability density for the measured phase
$\phi(H)$ is given by
\begin{equation}\label{eq:Px}
      P(\phi(H)|\vec{x}\>) = \frac1{2\pi} \, \frac{\sin^2[d^K (\phi(H)
  -\tilde\phi_{\vec{x}})/2]}{d^K\sin^2[(\phi(H)-\phi_{\vec{x}})/2]},
\end{equation}
where $\vec{x}$ is a string of $K$ base-$d$ digits.  Using the relation
$\lim_{\gamma\to\infty} [\sin^2(\gamma x)/\pi\gamma x^2] = \delta(x)$ one
easily checks that both results approach the limit of a $\delta$-function
$P(\phi(H)|\vec{x}\>) \to \delta (\phi(H)-\tilde\phi_{\vec{x}})$ when $K \to \infty$.

\begin{figure}[htbp]
\begin{center}
\includegraphics[width=8truecm]{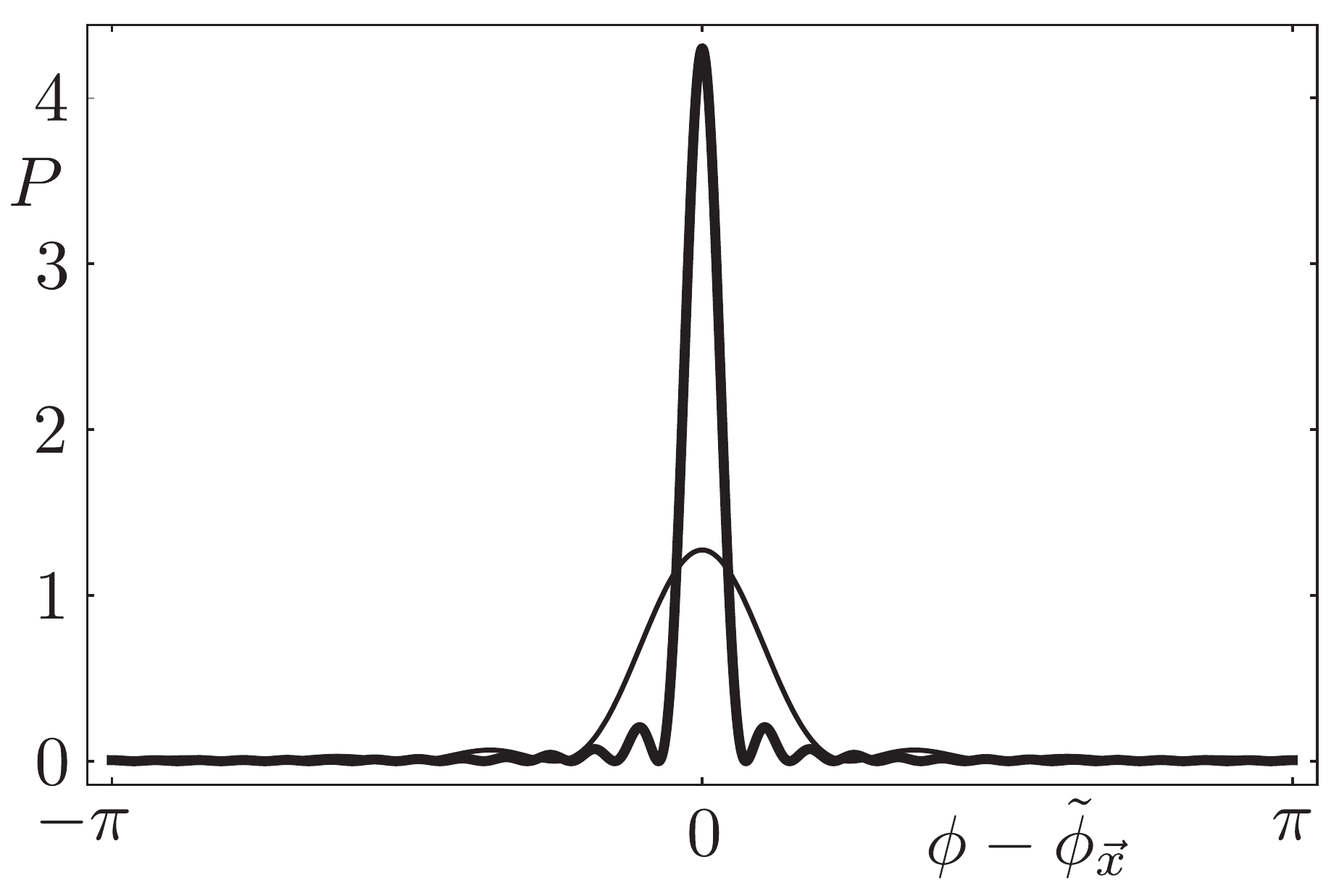}
\caption{The probability density plots $P(\phi-\tilde\phi_{\vec{x}})$ for the
$K=3$ step Fourier metrological procedure operated in the qutrit (thick line)
and qubit (thin line) regimes.}
\label{fig:density}
\end{center}
\end{figure}

Next, we discuss the impact of a false digit assignment on the measurement
outcome of the Fourier metrological scheme. This follows from the probability
density plot $P(\phi|\vec{x}\>) \equiv P(\delta\phi)$ evaluated as a function
of the estimation error $\delta\phi = \phi - \tilde\phi_{\vec{x}}$ which is
shown in Fig.~\ref{fig:density} for the qubit and qutrit based algorithms.

Both plots show a sharp central peak $\delta\phi \in[-2\pi/d^K,2\pi/d^K]$,
$d=2$ or $3$, and a number of decaying satellite peaks. These satellite peaks
derive from a wrong assignment of the binary $b_i$ or trinary $t_i$ digit
during the measurement run. The first satelites correspond to a false
assignment of the least significant digit in the first step of the procedure,
while the far weaker satellites further out correspond to assignment errors of
subsequent readouts. This analysis shows that the Fourier metrological
procedure is stable with respect to the assignment errors: the probability to
determine a false digit decreases for each next step of the procedure,
resulting in a confidence level that is highest for the most significant
digits and decreases for the measurement of the less significant digits. The
probability that the observed value of the phase $\phi$ lies within the region
of the central peak, and hence no error has been made in the assignment of
digits, is given by
\begin{equation}
   P\Bigl( \delta \phi \in \Bigl[ -\frac{2\pi}{d^K},
   \frac{2\pi}{d^K}\Bigr] \Bigr) \approx \frac1\pi\int\limits_{-\pi}^\pi dy
   \frac{\sin^2y}{y^2} \approx 0.903,
\end{equation}
where we have assumed a large $K$: the error probability saturates and does
not depend on the number of steps $K$ (or, equivalently, the number of
digits), manifesting the stability of the Fourier procedure.

\section{Comparing the qubit- and qutrit procedures}\label{sec:comp}

The posterior distribution functions for the phase $\phi(H)$ in the qutrit-
and qubit metrological procedures, see Eqs.\ (\ref{eq:qutritprob}) and
(\ref{eq:qubitprob}), allows us to compare the efficiency of the two schemes
quantitatively and reveal the advantage of using a higher-dimensional quantum
system for metrological purposes. Let the unknown magnetic field $H$ be
located somewhere within the continuous interval $H \in [0,H_0]$. Then, as
follows from the Eqs.\ (\ref{eq:Hbinary}) and (\ref{eq:Htrinary}), the field
scales for the qubit- and qutrit-based metrology are chosen as
$h_0^\mathrm{qb} = H_0/2$ and $h_0^\mathrm{qt} = H_0/3$ and the corresponding
minimal Ramsey delays are given by $\tau_0 = 2\pi\hbar/\mu H_0$. During the
$K$ steps, the qubit and qutrit metrological procedures learn about the
magnetic field to a precision $(\delta H)_\mathrm{qb} = H_0/2^K$ and $(\delta
H)_\mathrm{qt} = H_0/3^K$.  These $K$-step precision boundaries have to be
related to the amount of quantum resources required to achieve them. The
quantum resource exploited in our metrological procedures is the coherence of
the quantum devices, which can be quantified by the net coherence (or
phase-accumulation) time accumulated during the $K$ steps
\cite{Giovanetti:2004}. For each of the two protocols, this time is given by,
\begin{eqnarray}
  &&T_\mathrm{qb} = \tau_0 \sum_{k=0}^{K-1} 2^k
  \approx \frac{2\pi \hbar}{\mu H_0}\, 2^K, \\
  &&T_\mathrm{qt} = \tau_0 \sum_{k=0}^{K-1} 3^k
  \approx \frac{2\pi \hbar}{\mu H_0}\, \frac{3^K}2.
\end{eqnarray}
Expressing the $K$-step precisions $\delta H$ through the net coherence time,
\begin{eqnarray}
  (\delta H)_\mathrm{qb} = \frac{2\pi \hbar}{\mu}\, \frac1{T_\mathrm{qb}},
  \qquad (\delta H)_\mathrm{qt} = \frac{\pi \hbar}{\mu}\, \frac1{T_\mathrm{qt}},
  \label{eq:precision}
\end{eqnarray}
one notes that both procedures attain the Heisenberg limit, with the precision
$\delta H$ scaling as the inverse of the total coherence time, but the qutrit
procedure has a twice better prefactor. For the general qudit metrological
scheme, the combination of $(\delta H)_\mathrm{qd} = H_0/d^K$, $T_\mathrm{qd}
= \tau_0 d^K/(d-1)$, and $\tau_0 =2\pi\hbar/\mu H_0$ produces the
sensitivity,
\begin{equation}
   (\delta H)_\mathrm{qd} = \frac{2\pi\hbar}{\mu}\, \frac{1}{(d-1) T_\mathrm{qd}}.
   \label{eq:precision2}
\end{equation}

In practice, however, the phase accumulation time is the minimal time used in
a Ramsey experiment. Most of the overall measurement time is spent on the
measurement and re-initialization of the quantum devices.  Hence, in practice,
the speed of the metrological procedure is mostly defined by the number $K$ of
steps required to achieve a given precision. Here, the qutrit-based procedure
has a clear advantage: in order to achieve a relative precision $\delta H/H_0$
it requires approximately $K \sim -\log_3(\delta H/H_0)$ steps, that is
$\ln(2)/\ln(3) \approx 0.63$ fewer than the number of steps $K \sim
-\log_2(\delta H/H_0)$ needed for the qubit based scheme.

\section{Metrology with a transmon device} \label{sec:transmon}

The working principle of a transmon qubit as a magnetic-flux sensor has been
recently demonstrated in Ref.\ [\onlinecite{Danilin:2017}]; here, we describe
how the qutrit metrological protocol can be realized with a transmon device.
The superconducting transmon \cite{Koch:2007} involves a capacitively shunted
SQUID loop and constitutes an excellent candidate to implement qudit
metrological algorithms. Its dynamics is described by the Hamiltonian
\begin{equation}
    \hat{H} = 4E_{\rm\scriptscriptstyle C}(\hat{n}-n_g)^2
    - E_{\rm\scriptscriptstyle J}(\Phi) \cos(\hat\varphi),
      \label{eq:transmon_ham}
\end{equation}
where $\hat{n}$ is the number of Cooper pairs transmitted between
superconducting islands with a charging energy $E_{\rm \scriptscriptstyle C}$
and relative  phase $\hat\varphi$ and $n_g$ is the charge bias.  The Josephson
energy $E_{\rm\scriptscriptstyle J}(\Phi)$ of the SQUID loop depends on the
flux $\Phi$ through the loop,
\begin{equation}
   E_{\rm\scriptscriptstyle J}(\Phi) = E_{{\rm \scriptscriptstyle J}
     {\scriptscriptstyle \Sigma}}
   \sqrt{\cos^2\bigl( \pi {\Phi}/{\Phi_0}\bigr)
   + a^2 \sin^2\bigl( \pi {\Phi}/{\Phi_0}\bigr)}.
   \label{eq:jjenergy}
\end{equation}
Here, $E_{{\rm \scriptscriptstyle J} {\scriptscriptstyle \Sigma}} =
E_{\rm\scriptscriptstyle J1} + E_{\rm\scriptscriptstyle J2}$ is the total
energy of the two Josephson junctions in the SQUID loop, $a =
(E_{\rm\scriptscriptstyle J1}-E_{\rm\scriptscriptstyle J2})/E_{{\rm
\scriptscriptstyle J} {\scriptscriptstyle \Sigma}}$ is the junctions'
asymmetry, and $\Phi_0$ is the magnetic flux quantum.

The transmon atom is operated in the limit where $E_{\rm\scriptscriptstyle
J}/E_{\rm\scriptscriptstyle C} \sim 80 -200$. Its energy spectrum comprises a
discrete set of non-equidistant energy levels with positions that depend on
the magnetic flux $\Phi$ penetrating the SQUID loop of the device. Expanding
the $\cos(\hat\varphi)$ term in the Hamiltonian Eq.\ (\ref{eq:transmon_ham})
and treating the term quartic in $\hat\varphi$ as a perturbation, we obtain
(to leading order) the transmon's energy spectrum in the form
\begin{equation}
   E_n \approx \sqrt{8E_{\rm\scriptscriptstyle C} E_{\rm\scriptscriptstyle J}(\Phi)}
   \Bigl( n + \frac12 \Bigr) - E_{\rm\scriptscriptstyle J}(\Phi)
   -\frac{E_{\rm\scriptscriptstyle C}}{12}\bigl(6n^2+6n+3\bigr),
   \label{eq:spectrum}
\end{equation}
The nonlinearity of the spectrum $E_{n+1}-E_n = -E_{\rm\scriptscriptstyle
C}(n+1) + \sqrt{8E_{\rm\scriptscriptstyle C} E_{\rm\scriptscriptstyle
J}(\Phi)}$ allows for the individual addressing of the transmon's quantum
states through application of pulses of electromagnetic radiation with
specific frequencies. On the other hand, to leading order, the dependence of
the spectrum on the magnetic flux $\Phi$ is linear in $n$; while second-order
corrections $\propto E_{\rm\scriptscriptstyle C} (E_{\rm\scriptscriptstyle
J}/E_{\rm\scriptscriptstyle C})^{-1/2}$ do modify this result, these
corrections are small and we neglect them in the following.

In order to manipulate the first two excited states of our transmon atom, we
consider a two-tone rf-pulse that generates a time-varying electric potential
difference of the form
\begin{equation}
  V(t) = \Omega(t) \bigl( V_1 \cos(\omega_1 t) + V_2 \cos(\omega_2 t) \bigr),
\end{equation}
at the transmon's capacitor, where $\Omega(t)$ is the pulse envelope and
$V_{1(2)}$ are the amplitudes of the pulse components with tone frequencies
$\omega_{1(2)}$.  The transmon evolution under the pulse $V(t)$ then is
described by the Hamiltonian
\begin{eqnarray}
  \hat{H} = \sum_{n=0}^\infty E_n |n\rangle \langle n|
  + \bigl( \hbar g_{n,n+1}(t) |n\rangle \langle n\!+\!1| +
  \mathrm{h.c.} \bigr),
\end{eqnarray}
with the transition amplitudes $\hbar g_{n,n+1}(t) = 2\beta eV(t)$ $\langle
n|\hat{N}| n\!+\!1\rangle$; here, $\hat{N}$ is the number operator of Cooper
pairs transferred between the transmon's capacitor plates and $\beta$ is a
geometrical factor which quantifies the coupling between the transmon's
capacitor and the rf-field, see Ref.\ [\onlinecite{Koch:2007}].

Let the tone frequencies $\omega_1$ and $\omega_2$ be near the transition
frequencies $\omega_{01} = (E_1-E_0)/\hbar$ and $\omega_{12} =
(E_2-E_1)/\hbar$ of the first two pairs of levels, such that only the three
lowest energy levels are affected by the rf-field. We work in the rotating
frame (or interaction representation) with respect to the `free' Hamiltonian
$\hat{H}_0 = \hbar\omega_1 |1\rangle \langle 1| + \hbar(\omega_1+\omega_2)
|2\rangle \langle 2|$. A quantum state of this effective three-level system
can be represented as $|\Psi(t)\rangle = a_0(t) |0\rangle + a_1(t)|1\rangle
+a_2(t) |2\rangle$.  The time-dependent amplitudes $\vec{a}(t) =
[a_0(t),a_1(t),a_2(t)]$ obey the Schr\"odinger equation $i\hbar\partial_t
\vec{a}(t) = \hat{H}(t) \vec{a}(t)$, with the Hamiltonian assuming the
following form in the rotating wave approximation,
\begin{equation}
      \hat{H}(t) = \hbar \left[ \begin{array}{ccc}
      0& \Omega(t) \Delta_1& 0\\
      \!\! \Omega(t)\Delta_1& \omega_{01}\!-\omega_1& \Omega(t)\Delta_2\\
      0&\Omega(t)\Delta_2& \omega_{01}\!+\omega_{12}\!-\! \omega_1\!
      -\!\omega_2\!\!\end{array}\right],
      \label{eq:hamiltonian}
\end{equation}
where $\Delta_1 = \beta e V_1 \langle 0|\hat{N}|1\rangle/\hbar$ and $\Delta_2
= \beta e V_2 \langle 1|\hat{N}|2\rangle/\hbar$ are effective transition
amplitudes.

\subsection{Qutrit metrological protocol}\label{sec:protocol}

Our qutrit metrological scheme involves three steps that have to be
implemented with the help of proper manipulation signals $V(t)$.  In the first
step, we prepare the transmon in a balanced superposition of the form
\begin{equation}
   |\Psi_0\rangle \!=\! \frac1{\sqrt{3}} \bigl( e^{i\varphi_0} |0\rangle
   + e^{i\varphi_1}|1\rangle + e^{i\varphi_2} |2\rangle \bigr)
   \!\equiv \! \frac1{\sqrt{3}} \left[ \begin{array}{c} \!\!e^{i\varphi_1}\!\!
   \\ \!\!e^{i\varphi_2}\!\! \\ \!\!e^{i\varphi_3}\!\!\end{array} \right]\!.
\end{equation}
In the second step, the free evolution $\hat{U}(\Phi)$ of the transmon
generates the additional phase factors $|1\rangle \to e^{i\phi}|1\rangle$ and
$|2\rangle \to e^{i\phi^\prime} |2\rangle$ with $\phi = [\omega_{01}(\Phi) -
\omega_1]\tau$ and $\phi^\prime = [\omega_{01}(\Phi) + \omega_{12}(\Phi) -
\omega_1 -\omega_2]\tau$.  The remarkable property of the transmon is that its
level separations scale equally in magnetic flux, see Eq.\
(\ref{eq:spectrum}).  Starting from the reference magnetic flux $\Phi_c$, we
set the frequencies $\omega_1 = \omega_{01}(\Phi_c)$ and $\omega_2=
\omega_{12}(\Phi_c)$. Then, $\phi^\prime = 2\phi \equiv 2
[\omega_{01}(\Phi)-\omega_{01}(\Phi_c)]\tau$ and the state $|\Psi_0\rangle$
evolves to the new balanced state
\begin{equation}
      |\Psi_\phi\rangle = \frac1{\sqrt{3}} \left[ \begin{array}{c}
      \!\!e^{i\varphi_0}\!\!\\
      \!\!e^{i\varphi_1+i\phi}\!\!\\
      \!\!e^{i\varphi_2+2i\phi}\!\!
      \end{array} \right].
      \label{eq:phitau}
\end{equation}
Finally, in the third step of our qutrit metrological procedure, we need to
construct a unitary readout operator of the form
\begin{widetext}
\begin{equation}
      \hat{U}_r = \frac1{\sqrt{3}} \left[ \begin{array}{ccc}
      \!\! e^{i\chi_0}&0&0 \!\!\\
      \!\! 0&e^{i\chi_1}&0 \!\!\\
      \!\! 0&0&e^{i\chi_2} \!\!
      \end{array}\right]
      \left[ \begin{array}{ccc}
      \!\! 1&1&1 \!\!\\
      \!\! 1&e^{-2\pi i/3}&e^{+2\pi i/3} \!\!\\
      \!\! 1&e^{2\pi i/3}&e^{-2\pi i/3}
      \end{array}\right]
      \left[ \begin{array}{ccc}
      \!\! e^{-i\varphi_0}&0&0 \!\!\\
      \!\! 0&e^{-i\varphi_1}&0 \!\!\\
      \!\! 0&0&e^{-i\varphi_2} \!\!
      \end{array}\right],
      \label{eq:readout}
\end{equation}
\end{widetext}
that represents a generalized base-3 inverse Fourier transform.

For the specific situation where the accumulated phase $\phi$ can only assume
the three values $0, 2\pi/3$, and $4\pi/3 \leftrightarrow -2\pi/3$, this
measurement scheme is deterministic and the transmon will always be found in
one of the pure states $|0\rangle$, $|1\rangle$, or $|2\rangle$.  Indeed, for
$\phi=2\pi/3$ or $\phi=4\pi/3$, the state $|\Psi_0\rangle$ transforms into
states
\begin{equation}
      |\Psi_1\rangle = \frac1{\sqrt{3}} \left[ \begin{array}{c}
      \!\!e^{i\varphi_0}\!\!\\
      \!\!e^{i\varphi_1+i \frac{2\pi}{3}}\!\!\\
      \!\!e^{i\varphi_2+i \frac{4\pi}{3}}\!\!
      \end{array} \right],\>
      |\Psi_2\rangle = \frac1{\sqrt{3}} \left[ \begin{array}{c}
      \!\!e^{i\varphi_0}\!\!\\
      \!\!e^{i\varphi_1+i \frac{4\pi}{3}}\!\!\\
      \!\!e^{i\varphi_2+i \frac{2\pi}{3}}\!\!
      \end{array} \right],
\end{equation}
which together with $|\Psi_0\rangle$ form the (orthonormal) computational basis.
Then, the readout operation $\hat{U}_r$ provides a deterministic outcome,
\begin{equation}
   \hat{U}_r|\Psi_j\rangle = e^{i\chi_j}|j\rangle, \quad j=0,1,2.
\end{equation}

For an arbitrary accumulated phase, the scheme is probabilistic and provides
the probabilities
\begin{eqnarray}
    \label{eq:prob}
    P_j(\phi) &=& |\langle j| \hat{U}_r \hat{U}(\Phi) \hat{U}_p|0\rangle|^2 \\
	&=& \frac19 \bigl[ 1 + 2\cos\bigl(\phi(\Phi) - 2\pi\,j/3 \bigr)
	\bigr]^2,
	   \quad j = 0,1,2,
    \nonumber
\end{eqnarray}
to observe the transmon in the state $|j\rangle$. The possible phase values
$\phi$ then are divided into the three sectors $S_0=[-{\pi}/{3},{\pi}/{3}]$,
$S_1 = [\pi/3,\pi]$ and $S_2 = [\pi,{5\pi}/3]$ with the maximal probability
$P_j$ telling that $\phi \in S_j$.

\subsection{rf-pulses for metrological protocol}

In order to implement our metrological protocol, we have to find appropriate
rf-pulses $V_p(t)$ and $V_r(t)$ that prepare the transmon in a balanced state
and readout the state after its free evolution in the magnetic field to be
measured.  It turns out to be convenient to reverse the order and first find
the readout pulse.

Hence, our next goal is to find an rf-pulse $V_r(t)$ that generates a unitary
readout operation of the form (\ref{eq:readout}), a task that we tackle in
three steps: i) We show that any $3\times3$ unitary with  equal-modulus matrix
elements $|\hat{U}_{ij}| = {1}/{\sqrt{3}}$ is either of the form $\hat{U}_r$
or $\hat{U}_r^{-1}$. ii) We find the unitary associated with a rectangular
two-tone rf-pulse of finite duration $\tau_p$. iii) We determine the
constraints on the rf-pulse required for the readout action.

Starting with i), we consider an arbitrary $3\times3$ unitary matrix $\hat{U}$
with all matrix elements of modulus $1/\sqrt{3}$. After multiplication with
suitable diagonal phase matrices from the left and right, we can arrive at the
form
\begin{equation}
      \hat{U} \to \hat{U}' = \frac1{\sqrt{3}} \left[ \begin{array}{ccc}
      1&1&1\\
      1&e^{i\alpha_1}&e^{i\beta_1}\\
      1&e^{i\alpha_2}&e^{i\beta_2} \end{array}\right],
\end{equation}
where the remaining four phases have to satisfy the orthogonality constraints
between columns, $1 + e^{i\alpha_1}+e^{i\alpha_2} = 0$, $1 + e^{i\beta_1}
+e^{i\beta_2} =0$ and $1 +e^{i(\alpha_1-\beta_1)}+e^{i(\alpha_2 - \beta_2)} =
0$. It follows that the solutions of these constraints define either the
Fourier transform $\hat{U}' = \hat{F}_3$ or its inverse $\hat{U}' =
\hat{F}_3^{-1}$, see Eq.\ \eqref{eq:invFourier}, that proves our statement.

Next, in step ii), we consider a two-tone rf-pulse with a rectangular shape of
duration $\tau_p$ and frequencies $\omega_1 = \omega_{01}(\Phi_c) -
2\delta\omega$ and $\omega_2 = \omega_{12}(\Phi_c)+2\delta\omega$. Such a
pulse generates a unitary rotation of the qutrit $\hat{U} =
\exp(-i\hat{H}\tau_p/\hbar) \equiv \exp(-i\hat{K})$, where
\begin{equation}
   \hat{K} = \left[ \begin{array}{ccc}
   0&\Delta_1&0\\
   \Delta_1&2\epsilon&\Delta_2\\
   0&\Delta_2&0 \end{array} \right],
   \quad \epsilon = \delta\omega \,\tau_p,
\end{equation}
and $\Delta_{1,2}$ are effective transition amplitudes, see Eq.\
\eqref{eq:hamiltonian}. The resulting unitary transformation has the form,
\begin{widetext}
\begin{equation}
   \hat{U} = \left[ \begin{array}{ccc}
   \frac{\Delta_2^2}{\Delta_1^2+\Delta_2^2}&0&
          -\frac{\Delta_1\Delta_2}{\Delta_1^2+\Delta_2^2}\\
   0&0&0\\
   -\frac{\Delta_1\Delta_2}{\Delta_1^2 +\Delta_2^2}&0&
           \frac{\Delta_1^2}{\Delta_1^2+\Delta_2^2}
   \end{array}\right]
   +e^{-i\epsilon} \cos(\xi) \left[ \begin{array}{ccc}
   \frac{\Delta_1^2}{\Delta_1^2+\Delta_2^2}&0&
           \frac{\Delta_1\Delta_2}{\Delta_1^2 +\Delta_2^2}\\
   0&1&0\\
   \frac{\Delta_1\Delta_2}{\Delta_1^2 +\Delta_2^2}&0&
           \frac{\Delta_2^2}{\Delta_1^2+\Delta_2^2}
   \end{array}\right]
   +\frac{ie^{-i\epsilon}\sin(\xi)}{\xi} \left[ \begin{array}{ccc}
    \frac{\epsilon\Delta_1^2}{\Delta_1^2+\Delta_2^2}&-\Delta_1&
           \frac{\epsilon\Delta_1\Delta_2}{\Delta_1^2 +\Delta_2^2}\\
   -\Delta_1&-\epsilon&-\Delta_2\\
   \frac{\epsilon\Delta_1\Delta_2}{\Delta_1^2 +\Delta_2^2}&-\Delta_2&
           \frac{\epsilon\Delta_2^2}{\Delta_1^2+\Delta_2^2}
   \end{array}\right],
\end{equation}
\end{widetext}
where $\xi = \sqrt{\epsilon^2+ \Delta_1^2 +\Delta_2^2}$.

Let us then, iii), determine the parameters that generate the readout matrix
$\hat{U}_r$. Following i), we have to require that $|[U_r]_{ij}|^2 = 1/3$. The
conditions $|U_{12}|^2 = |U_{21}|^2 =1/3$ and $|U_{23}|^2 = |U_{32}|^2 = 1/3$
imply that
\begin{equation}
   \frac{\sin^2(\xi)}{\xi^2}\, \Delta_1^2 = \frac13, \quad
   \frac{\sin^2(\xi)}{\xi^2}\, \Delta_2^2 = \frac13,
\end{equation}
that gives $\Delta_1^2 = \Delta_2^2 \equiv \Delta^2$. Accounting for the
remaining conditions $|U_{11}|^2=|U_{13}|^2=1/3$ and using the relation
$2\Delta^2 = \xi^2-\epsilon^2$, we arrive at the following system of
trans\-zendental equations
\begin{eqnarray}
   &&\epsilon^2 = \xi^2 \Bigl( 1- \frac{2}{3\sin^2(\xi)} \Bigr),
   \label{eq:sys1}
   \\
   && \cos(\epsilon)\cos(\xi)+ \frac{\epsilon}{\xi} \sin(\epsilon)\sin(\xi) = 0.
   \label{eq:sys2}
\end{eqnarray}
Among its solutions, we choose the one with the minimal $\epsilon$ as it
corresponds to the shortest rf-pulse for a given detuning $\delta\omega$ and
find the numerical values $\epsilon_0 \approx 0.8525$, $\xi_0 \approx 2.0205$,
and hence $\Delta_0 \approx 1.2953$. We note, that the system of Eqs.\
(\ref{eq:sys1}) and (\ref{eq:sys2}) remains unchanged under a sign-change of
the parameters $\epsilon$, $\Delta_1$, and $\Delta_2$ characterizing the
Hamiltonian.

Let us consider the specific solution with $\epsilon = -\epsilon_0$ and
$\Delta_1 = \Delta_2 = \Delta_0$. The corresponding pulse then generates the
readout unitary transformation
\begin{eqnarray}
   &&\hat{U}_r = \frac1{\sqrt{3}} \left[ \begin{array}{ccc}
   e^{-i\frac{\pi}{6}}&-ie^{i\epsilon_0}&e^{-i\frac{5\pi}{6}}\\
     -ie^{i\epsilon_0}&ie^{i2\epsilon_0}&-ie^{i\epsilon_0}\\
       e^{-i\frac{5\pi}{6}}&-ie^{i\epsilon_0}&e^{-i\frac{\pi}{6}}
   \end{array}\right]
   \label{eq:Ur} \\
   &&\equiv \left[ \begin{array}{ccc}
      1&0&0\\
        0&e^{i\epsilon_0+i\frac{5\pi}{6}}&0\\
          0&0&e^{i\frac{4\pi}{3}}
   \end{array}\right]
   \hat{F}_3^{-1} \left[ \begin{array}{ccc}
      e^{-i\frac{\pi}{6}}&0&0\\
        0&-ie^{i\epsilon_0}&0\\
          0&0&e^{-i\frac{5\pi}{6}}
   \end{array}\right],
   \nonumber
\end{eqnarray}
that provides the desired inverse generalized Fourier transform.

As a preparation operator $\hat{U}_p$, we choose a unitary rotation generated
by a rf-pulse with $\epsilon = +\epsilon_0$ and $\Delta_1 = \Delta_2 =
-\Delta_0$, i.e., the preparation pulse generates the inverse of the readout
operator, $\hat{U}_p = \hat{U}_r^\dagger$.

The sign change in $\epsilon$ is trivially realized by inverting the detuning
$\delta\omega \to - \delta\omega$. In order to change the sign (or
more generally the phase) of the effective transition amplitudes
$\Delta_{1,2}$, one can proceed with an appropriate modulation of the voltage
signal $V(t)$. Making use of a standard IQ (In-phase and Quadrature)-mixing
scheme, an incoming high-frequency signal $\cos(\omega_{\rm \scriptscriptstyle
LO}t)$ with the (local oscillator) frequency $\omega_{\rm\scriptscriptstyle
LO} = \frac12 \bigl[ \omega_{01}(\Phi_c) + \omega_{12}(\Phi_c)\bigr]$ is first
physically split (and partly phase shifted) into two separate signals
$\cos(\omega_{\rm \scriptscriptstyle LO}t) \to \frac12 \bigl[\cos(\omega_{\rm
\scriptscriptstyle LO}t) + \sin(\omega_{\rm \scriptscriptstyle LO}t)\bigr]$.
These are independently mixed with the intermediate-frequency signals
$A_1\Omega(t) \cos(\omega_{\rm \scriptscriptstyle IF}t+\varphi)$ and $A_2
\Omega(t) \sin(\omega_{\rm \scriptscriptstyle IF}t +\varphi)$ generated by an
arbitrary-waveform generator.  Finally, the signals are recombined and the
resulting output signal sent to the transmon is given by
\begin{eqnarray}
   &&V(t) = \frac{\Omega(t)}4 \Bigl[(A_1-A_2)
   \cos\bigl[ (\omega_{\rm\scriptscriptstyle LO}
              +\omega_{\rm\scriptscriptstyle IF})t+\varphi\bigr]
      \\
   &&\qquad\qquad\quad + (A_1+A_2)
   \cos\bigl[(\omega_{\rm\scriptscriptstyle LO}
             -\omega_{\rm\scriptscriptstyle IF})t-\varphi\bigr]\Bigr]. \quad
      \nonumber
\end{eqnarray}
Choosing amplitudes $A_1 = 2(V_1+V_2)$, $A_2 = 2 (V_2-V_1)$, the frequency
$\omega_{\rm\scriptscriptstyle IF} = \frac12 \bigl[\omega_{01} (\Phi_c) -
\omega_{12}(\Phi_c) \bigr] - 2\delta\omega$, and the phase $\varphi = 0$, one
can generate the readout pulse. On the other hand, choosing the frequency
$\omega_{\rm\scriptscriptstyle IF} = \frac12 \bigl[\omega_{01}(\Phi_c) -
\omega_{12}(\Phi_c) \bigr] + 2 \delta \omega$ and the phase $\varphi = \pi$
together with the sign inversion of the detuning $\delta\omega$ reverses the
sign of all three parameters $\epsilon$, $\Delta_1$, and $\Delta_2$ and hence
produces the preparation pulse.

\subsection{Optimizing the transmon sensitivity}

As follows from Eq.~(\ref{eq:precision2}), the measurement precision that can
be attained by the qudit metrological protocol depends on two factors, the
magnetic moment $\mu$ of the transmon device and the longest phase coherence
time $d^{K-1}\tau_0$ required for the longest run of the metrological
protocol. The magnetic moment of the transmon can be obtained via the
curvature of its transition frequency,
\begin{equation}
   \mu = \hbar A \frac{\partial \omega_{01}(\Phi_c)}{\partial \Phi},
\end{equation}
where $A$ is the area of the SQUID loop. Indeed, the relative phase $\phi =
\bigl[ \omega_{01}(\Phi) - \omega_{01}(\Phi_c) \bigr]\tau$ accumulated by the
transmon's wavefunction is given by,
\begin{eqnarray}
   \phi \approx \tau \frac{\partial \omega_{01}(\Phi_c)}{\partial \Phi}
   (\Phi - \Phi_c) = \frac{\mu \delta H \tau}{\hbar},
\end{eqnarray}
where $\delta H = H - H_c$ is the magnetic field measured relative to the
reference magnetic field $H_c$, $\Phi_c = A H_c$. In order to attain a better
sensitivity, one has to deviate from the 'sweet spot', the upper maximum of
the transmon spectrum $\omega_{01}(\Phi)$ where $\mu = 0$, and take the device
to a (locally) linear regime with a lower transition frequency. In the limit
of an almost symmetric Josephson junction loop with $a \to 0$, the maximal
value of the transmon's magnetic moment is given by (see Eqs.\
(\ref{eq:spectrum}) and (\ref{eq:jjenergy}))
\begin{equation}\label{eq:mu}
   \mu = \pi \frac{A}{\Phi_0} \, \sqrt{\frac{8E_{\rm\scriptscriptstyle C}
   E_{{\rm \scriptscriptstyle J}
     {\scriptscriptstyle \Sigma}}}{a}},
\end{equation}
which occurs near the bottom of the transmon spectrum where $\tan^2(\pi
\Phi_c/\Phi_0) = 1/a$. The result Eq.\ \eqref{eq:mu} applies to the transmon
limit $E_{\rm \scriptscriptstyle J} \gg E_{\rm\scriptscriptstyle C}$; for a
symmetric device $a \to 0$, the largest moment $\mu$ appears near the point of
maximal frustration $\Phi_c = \Phi_0/2$ where $E_{\rm \scriptscriptstyle J}$
becomes small and the approximation breaks down.

In our discussion above, we have implicitly assumed that the entire
preparation--phase-accumulation--readout sequence involves a total time that
is much below the coherence- ($T_2$) and relaxation ($T_1$) times of the
transmon device. In a realistic situation, when operating the transmon away
form the 'sweet spot' the $T_2$-time gets reduced and so is the number $K$ of
available steps for the metrological procedure.  For a given qudit coherence
time $T_2$, the delay time of the longest Ramsey sequence cannot exceed the $T_2$
time. Hence, the maximum number of steps in the Fourier procedure is limited
by the condition $\tau_0 d^{K-1} = T_2$ that gives $K = 1 +
\log_d(T_2/\tau_0)$ steps, where $\tau_0$ is the minimum duration of a Ramsey
sequence. Thus, the total amount of coherence time spent for the signal
sensing is given by $T_\mathrm{qd} = \tau_0 \sum_{k=0}^{K-1} d^k = \tau_0
(d^{K-1}-1)/(d-1) \approx [d/(d-1)] T_2$ for $K \gg 1$.  Then, according to
the Eq.\ (\ref{eq:precision2}), the best attainable field-resolution can be
estimated as
\begin{equation}
      [\delta H]_{T_2} \approx \frac{2\pi \hbar}{\mu\, d\, T_2}.
\end{equation}
Hence, we have to optimize the product $\mu(\Phi_c)T_2(\Phi_c)$ for the best
flux bias $\Phi_c$, compromising between two opposing trends, a magnetic
moment $\mu(\Phi_c)$ increasing and the coherence time $T_2(\Phi_c)$
decreasing away from the sweet spot. The magnetic moment of a transmon atom
can attain a value of $10^5\mu_{\rm \scriptscriptstyle B}$, see Ref.\
[\onlinecite{Danilin:2017}].  Assuming a coherence time $T_2 \sim 1~\mu$s, one
can estimate that a magnetic-field precision of order $\delta H \sim 0.1$ nT
can be achieved.

A further improvement of the field resolution is possible only within a
standard statistical measurement scheme\cite{Sekatski:2017} by repeating the
longest Ramsey measurement with $\tau \sim T_2$ a large number $N\gg 1$ of
times.  This leads to a standard scaling of the further field resolution with
time duration $t$ of the experiment: with $N = t/T_2$ we obtain a precision
\begin{equation}
    [\delta H]_{t\gg T_2} \approx \frac{2\pi \hbar}{\mu\, d\, T_2\sqrt{t/T_2}}
    = \frac{2\pi \hbar}{\mu\, d\, \sqrt{T_2 t}}.
    \label{eq:standard}
\end{equation}
Therefore, the Heisenberg scaling is limited to measurement times shorter then
the coherence time $T_2$ of the device, with the standard quantum limit
restored for larger measurement times.  The standard scaling Eq.\
(\ref{eq:standard}) can be achieved by a conventional scheme where one always
measures the transmon state at the longest possible delay $T_2$ of the Ramsey
sequence. However, such a conventional scheme has a limited measurement range
$\Delta H$ for the field $H$ that results from the $2\pi$-periodicity of the
accumulated phase $\phi \sim \mu \Delta H T_2/\hbar$, i.e., $\Delta H \sim
2\pi\hbar/{\mu T_2}$. In contrast, the quantum procedure does not suffer from
this limitation: its measurement range is defined only by the minimal time
duration $\tau_0$ of the Ramsey sequence that is limited in practice by the
time duration of the controlling rf-pulses.

Making use of the above ideas in an experiment, a further practical
restriction has to be considered besides the finite coherence time of the
transmon device. First, the total duration of the experiment has to include
the additional time spent for the measurement and reset of the transmon state.
In fact, in the sensing experiment of Ref.\ \onlinecite{Danilin:2017} using
the transmon qubit-mode, most of the time of a single Ramsey measurement
$T_\mathrm{rep}$ has been spent on the measurement and reset of the qubit
$T_\mathrm{rep} \gg T_2$. In this limit, the long-time sensitivity of the
sensor is reduced and given by
\begin{equation}
   [\delta H]_{t\gg T_\mathrm{rep}\gg T_2} \approx \frac{2\pi \hbar}
   {\mu\, d\, T_2\, \sqrt{t/T_\mathrm{rep}}}.
   \label{eq:standard_rep}
\end{equation}
Second, the higher-excited states of the transmon have larger dipole matrix
elements and hence are more sensitive to the external electromagnetic
environment\cite{Koch:2007}. Therefore, a transmon atom has specific $T_1$ and
$T_2$ times for each excited state and the number of levels which can be used
is naturally limited by the coherence time of the highest-energy state
involved. In practice, one can start the Fourier metrological procedure in a
qubit measurement mode at Ramsey delays corresponding to the largest $T_2$
time belonging to the first excited level and then continue in a qutrit mode
when the Ramsey delays have dropped below the $T_2$ time of the second excited
level. Finally, operating a transmon atom in a qudit regime requires a more
involved characterization of its spectrum and calibration of the corresponding
rf-control pulses. At the same time, operating a transmon in the qutrit regime
is a well established experimental procedure
\cite{Abdulmalikov:2013,Kumar:2016}, what motivates work directed at the
experimental implementation of our qutrit metrological procedure.

\section{Summary and conclusion}\label{sec:conclusion}

We have presented a variant of the standard quantum Fourier metrological
procedure that replaces the usual qubit elements by qutrits and, more general,
by qudits. While, all of these algorithms exploit phase coherence as their
quantum resource, allowing them to reach the Heisenberg precision scaling, the
use of higher-dimensional Hilbert spaces in the qutrit and qudit versions
improves on the prefactor of this scaling. Even more, going to qudit devices
reduces the number of iteration steps in the Fourier procedure, thus providing
a marked improvement of the measurement algorithm. As a specific example, we
have discussed the use of a superconducting transmon device operated in the
qutrit mode that serves as an ideal resource for the measurement of dc and
low-frequency magnetic fields, a consequence of the linear field-dependence of
the transmon's spectrum. It turns out, that a simple two-tone rf voltage
signal in combination with a standard IQ-mixing scheme suffices to produce the
appropriate preparation and readout pulses for the Fourier metrological
algorithm.

The scheme presented in this paper relies on the assumption that the longest
measurement providing the least relevant but precision-limiting digit can be
performed within the coherence- or $T_2$-time of the transmon device; a
further increase in precision proceeds via a conventional measurement
procedure and follows the standard (shot-noise- or quantum-limited) scaling in
precision. Furthermore, given a finite coherence time $T_2$, the algorithm can
be further optimized to deal with this situation:  Besides properly tuning the
qutrit's reference flux or working point $\Phi_c$ as discussed above, other
elements of the algorithm can be improved.  E.g., a finite $T_2$-time may
require more than a single Ramsey measurement at each time delay $\tau_k$,
$k=1,\dots K$, what modifies the probability Eq.\ \eqref{eq:Px} to arrive at
the correct sequence $\vec{x}$ of digits.  In addition, the time delays
$\tau_k$ appearing in the metrological protocol are subject to optimization,
i.e., they must be chosen differently from the ideal case. The situation with
finite $T_1$- and $T_2$-times then requires a separate study that will be the
topic of a future analysis.

We acknowledge discussions with Pertti Hakonen and Vladimir Manucharyan.  The
research was supported by Government of the Russian Federation (Agreement
05.Y09.21.0018), by the RFBR Grants No. 17- 02-00396A and 18-02-00642A,
Foundation for the Advancement of Theoretical Physics ”BASIS”, the Ministry of
Education and Science of the Russian Federation 16.7162.2017/8.9 (A.V.L.,
A.R.S and V.V.Z. in part related to Section IV), the Swiss National Foundation
via the National Centre of Competence in Research in Quantum Science and
Technology (NCCR QSIT), the Pauli Center for Theoretical Physics,
Academy of Finland Centers of Excellence
"Low Temperature Quantum Phenomena and Devices" (project 250280) and
"Quantum Technology Finland (QTF)" (project 312296), and the Center
for Quantum Engineering at Aalto University. V.Z. and A.S. acknowledge the support from the 5-top 100 programm via the Laboratory of quantum information theory (MIPT).

\end{document}